\documentclass[a4paper]{jpconf}
\usepackage{graphicx}
\def\gsim{\raise0.3ex\hbox{$\;>$\kern-0.75em\raise-1.1ex\hbox{$\sim\;$}}}
\begin{document}

\title{Boltzmann equation with a nonlocal collision term and the
resultant dissipative fluid dynamics}

\author{Amaresh Jaiswal, Rajeev S Bhalerao and Subrata Pal}

\address{Tata Institute of Fundamental Research,
Homi Bhabha Road, Mumbai 400005, India}

\ead{amaresh@tifr.res.in, bhalerao@tifr.res.in, spal@tifr.res.in}

\begin{abstract}
Starting with the relativistic Boltzmann equation where the collision
term was generalized to include gradients of the phase-space
distribution function, we recently presented a new derivation of the
equations for the relativistic dissipative fluid dynamics. We compared
them with the corresponding equations obtained in the standard
Israel-Stewart and related approaches. Our method generates all the
second-order terms that are allowed by symmetry, some of which have
been missed by the traditional approaches, and the coefficients of
other terms are altered. The first-order or Navier-Stokes equation
too receives a small correction. Here we outline this work for the
general audience.

\end{abstract}

\section{Introduction}

The kinetic or transport theory of gases is a microscopic description
in the sense that detailed knowledge of the motion of the constituents
is required. Fluid dynamics (also sloppily called hydrodynamics) is an
effective (macroscopic) theory that describes the slow,
long-wavelength motion of a fluid close to local thermal
equilibrium. No knowledge of the motion of the constituents is
required to describe observable phenomena. Quantitatively, if $l$
denotes the mean free path, $\tau$ the mean free time, $k$ the wave
number, and $\omega$ the frequency, then $kl \ll 1,~ \omega \tau \ll
1$ is the hydro regime, $kl \simeq 1,~ \omega \tau \simeq 1$ the
kinetic regime, and $kl \gg 1,~ \omega \tau \gg 1$ the free-particle
regime.

Hydrodynamic equations are a set of coupled partial differential
equations for number density $n$, energy density $\epsilon$, pressure
$P$, hydrodynamic four-velocity $u^\mu$, and dissipative fluxes such
as bulk viscosity $\Pi$, heat current $n^\mu$, and shear stress tensor
$\pi^{\mu\nu}$. In addition, the equation of state (EoS) needs to be
supplied. Hydrodynamics is a powerful technique: Given the initial
conditions and the EoS, it predicts the evolution of the matter. Its
limitation is that it is applicable at or near (local) thermal
equilibrium only.

Hydrodynamics finds applications in cosmology, astrophysics,
high-energy nuclear physics, etc. In relativistic heavy-ion
collisions, it is used to calculate the multiplicity and transverse
momentum spectra of hadrons, anisotropic flows and femtoscopic
radii. Energy density or temperature profiles resulting from
the hydrodynamic evolution are needed in the calculations of jet
quenching, $J/\psi$ melting, thermal photon and dilepton
productions, etc. Thus hydro plays a central role in modeling
relativistic heavy-ion collisions.

\begin{figure}[t]                                                            
\begin{minipage}{18.pc}
\includegraphics[width=18.pc]{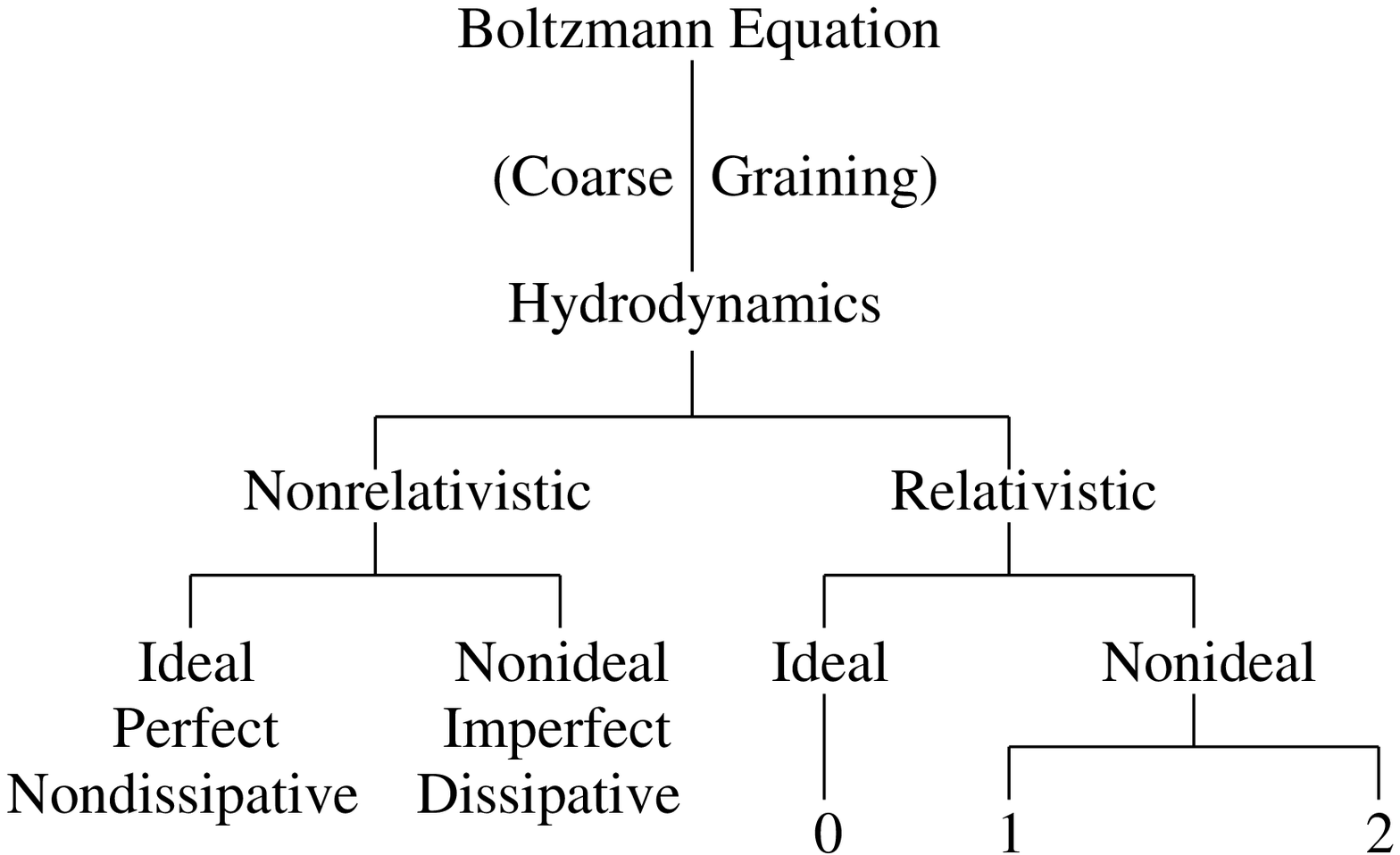}
\caption{\label{cgkt}Coarse-graining of kinetic theory.}
\end{minipage}\hspace{4.5pc}%
\begin{minipage}{15pc}
\includegraphics[width=13.2pc]{cartoon.eps}
\caption{\label{flel}Collisions $kk' \to pp'$ and $pp' \to kk'$ occurring at
  points $x^\mu$ and $x^\mu-\xi^\mu$ within an infinitesimal fluid
  element of size $dR$, containing a large number of particles
  represented by dots.}
\end{minipage}
\end{figure}

Hydrodynamics is formulated as an order-by-order expansion in
gradients of $u^\mu$, the ideal hydrodynamics being of the zeroth
order. The zeroth-, first-, and second-order equations are named after
Euler, Navier-Stokes, and Burnett, respectively, in the
non-relativistic case (Fig. \ref{cgkt}). The relativistic Navier-Stokes
equations are parabolic in nature and exhibit acausal behaviour, which
was rectified in the (second-order) Israel-Stewart (IS) theory
\cite{Israel:1979wp}. 
The formulation of the relativistic dissipative second-order
hydrodynamics (``2'' in Fig. \ref{cgkt}) is currently under intense
investigation
\cite{Baier:2006um,Baier:2007ix,Bhattacharyya:2008jc,Natsuume:2007ty,Bhalerao:2007ek,El:2008yy,El:2009vj,Denicol:2010xn,Denicol:2012cn}.

Hydrodynamics has traditionally been derived either from entropy
considerations (i.e., the generalized second law of thermodynamics) or
by taking the second moment of the Boltzmann equation; for a review,
see \cite{Romatschke:2009im}. 
The former approach captures a subset of the terms allowed
by symmetry \cite{Bhattacharyya:2012nq} 
in the evolution equations for the dissipative
quantities. The latter approach captures some more terms but not
all. The question `why do the traditional approaches not generate 
{\it all} the allowed terms', was the main motivation of our work 
\cite{Jaiswal:2012qm}.

\section{Present work}

In \cite{Jaiswal:2012qm} 
we presented a new derivation of the dissipative hydrodynamic
equations within kinetic theory but using a nonlocal collision term in
the Boltzmann equation. We obtained all the second-order terms that
are allowed by symmetry \cite{Bhattacharyya:2012nq} 
 and showed that the coefficients of the
existing terms in the widely used traditional IS equations were
altered. These modifications do have a rather strong influence on the
evolution of the viscous medium as we demonstrated in the case of
one-dimensional scaling expansion.

It is important to recall at the outset that an infinitesimal volume
element in the fluid is always supposed to be large compared with the
mean interparticle spacing and hence contains ``a very great'' number of
particles \cite{Landau};
 see also Fig. \ref{flel}. In kinetic theory, the
single-particle phase-space distribution function $f(x,p)$ is assumed
to vary slowly over space-time, i.e., it changes negligibly over the
range of interparticle interaction \cite{deGroot}, which in fact is
much smaller than the mean interparticle spacing. It is important to
keep this hierarchy of length scales in mind.

It is also interesting to recall Israel and Stewart's classic
paper \cite{Israel:1979wp}
 where they list the properties of the collision term
$C[f]$. We quote: ``We require only the following general
properties: (i) $C[f]$ is a purely local function or functional of
$f$, independent of $\partial_\mu f$. (ii) The form of $C[f]$ is
consistent with conservation of 4-momentum and number of particles at
collisions. (iii) $C[f]$ yields a non-negative expression for the
entropy production .... These requirements are of course met by
Boltzmann's ansatz for 2-particles collisions, and, indeed, one may
hope that they hold somewhat more generally, although the locality
assumption (i) is a powerful restriction.'' Note the words {\it hope}
and {\it assumption}. The locality assumption is questionable and we
relaxed it on the length scale of $dR$
\cite{Jaiswal:2012qm}.
Despite the long history of the Boltzmann
equation, a collision term containing $\partial_\mu f$, to
our knowledge, has never been used to derive hydrodynamic
equations. Such a collision term brings about a change not only in
hydrodynamics but also in the kinetic theory.

Our starting point is the relativistic Boltzmann equation with
the modified collision term:
\begin{equation}\label{MBE}
p^\mu \partial_\mu f = C_m[f] 
=  C[f] + \partial_\mu(A^\mu f) + \partial_\mu\partial_\nu(B^{\mu\nu}f) + \cdots,
\end{equation}
where $A^\mu$ and $B^{\mu\nu}$ depend on the type of the
collisions ($2 \leftrightarrow 2,~ 2 \leftrightarrow 3, \ldots$).

For instance, for $2 \leftrightarrow 2$ 
elastic collisions,
\begin{equation}\label{coll}
C[f]= \frac{1}{2} \int dp' dk \ dk' \  W_{pp' \to kk'}
\ (f_k f_{k'} \tilde f_p \tilde f_{p'}
- f_p f_{p'} \tilde f_k \tilde f_{k'}),
\end{equation}
where $ W_{pp' \to kk'}$ is the transition rate, $f_p
\equiv f(x,p)$ and $\tilde f_p \equiv 1-r f(x,p)$ with $r = 1,-1,0$
for Fermi, Bose, and Boltzmann gas, and $dp = g d{\bf p}/[(2 \pi)^3
\sqrt{{\bf p}^2+m^2}]$, $g$ and $m$ being the degeneracy factor and
particle rest mass. The first and second terms in Eq. (\ref{coll})
refer to the gain and loss processes $kk' \to pp'$ and $pp' \to kk'$,
respectively, occurring {\it anywhere} in the infinitesimal fluid element 
located at
the space-time point $x^\mu$. 
These processes have traditionally been assumed to occur at the
same point $x^\mu$ with an underlying assumption that
$f(x,p)$ is constant not only over the range of interparticle
interaction but also over the entire fluid element of
size $dR$. Boltzmann equation together with this crucial assumption
has been used to derive the standard second-order dissipative
hydrodynamic equations
\cite{Israel:1979wp,Denicol:2010xn,Romatschke:2009im}. 
We, however, emphasize that the variation of $f(x,p)$ over the
span of the fluid element may not be negligible,
and hence the space-time points at which the above two kinds of 
processes occur should be separated by an interval $|\xi^\mu| \leq dR$
within the volume $d^4R$ (Fig. \ref{flel}). 
The large number of
particles within $d^4R$ collide among themselves with various
separations $\xi^\mu$. Of course, the points $(x^\mu-\xi^\mu)$ must
lie within the past light-cone of the point $x^\mu$ (i.e., $\xi^2 > 0$
and $\xi^0>0$) to ensure that the evolution of $f(x,p)$ in
Eq. (\ref{MBE}) does not violate causality. With this realistic
viewpoint, the second term in Eq. (\ref{coll}) involves
$f(x-\xi,p)f(x-\xi,p')\tilde f(x-\xi,k)\tilde f(x-\xi,k')$, which on
Taylor expansion at $x^\mu$ up to second order in $\xi^\mu$, results
in the modified Boltzmann equation (\ref{MBE}) with
\begin{equation}\label{coeff1}
A^\mu = \frac{1}{2} \int dp' dk \ dk' \ \xi^\mu W_{pp' \to kk'}
f_{p'} \tilde f_k \tilde f_{k'},~~{\rm and}~~
B^{\mu\nu} = -\frac{1}{4} \int dp' dk \ dk' \ \xi^\mu \xi^\nu W_{pp' \to kk'}
f_{p'} \tilde f_k \tilde f_{k'}.
\end{equation}

In general, for all collision types ($2 \leftrightarrow 2,~ 2
\leftrightarrow 3, \ldots$), the momentum dependence of the
coefficients $A^\mu$ and $B^{\mu\nu}$ can be made explicit by
expressing them in terms of the available tensors $p^\mu$ and the
metric $g^{\mu\nu}$ as $A^\mu = a(x) p^\mu$
and $B^{\mu\nu}= b_1(x) g^{\mu\nu} + b_2(x) p^\mu p^\nu$. 
Equation
(\ref{MBE}) with this $A^\mu$ and $B^{\mu\nu}$
forms the basis of our derivation of the second-order
dissipative hydrodynamics. 
Arguments in the previous paragraph were meant only to provide a
physical motivation for the mathematical form of $C_m[f]$ in
Eq. (\ref{MBE}).

Conservation of current, $\partial_\mu N^\mu=0$ and energy-momentum tensor, $\partial_\mu
T^{\mu\nu} =0$, yield the fundamental evolution equations for $n$,
$\epsilon$ and $u^\mu$
\begin{eqnarray}\label{evol}
Dn+n\partial_\mu u^\mu + \partial_\mu n^\mu &=& 0, \nonumber \\
D\epsilon + (\epsilon+P+\Pi)\partial_\mu u^\mu - \pi^{\mu\nu}\nabla_{(\mu} u_{\nu)} &=& 0,  \nonumber\\
(\epsilon+P+\Pi)D u^\alpha - \nabla^\alpha (P+\Pi) + \Delta^\alpha_\nu \partial_\mu \pi^{\mu\nu}  &=& 0.
\end{eqnarray}
Equations (\ref{evol}) together with the EoS constitute 
six equations in fifteen unknowns. How to derive the extra nine equations
that would give us a closed set of equations? Boltzmann equation
provides a way: The requirement of the
conservation of energy-momentum and current
implies vanishing zeroth and first moments of the
collision term $C_m[f]$ in Eq. (\ref{MBE}), i.e., $\int dp \ C_m[f] =
0 = \int dp \ p^\mu C_m[f]$ at each order in $\xi^\mu$. 
In order to obtain the evolution equations for the dissipative
quantities, we follow the IS approach \cite{Israel:1979wp} and
consider the second moment of the modified Boltzmann equation
(\ref{MBE})
\begin{equation}\label{BE2}
\int dp \  p^\alpha p^\beta p^\gamma \partial_\gamma f 
= \int dp \ p^\alpha p^\beta \big[ C[f] + p^\gamma \partial_\gamma(af) 
+ \partial^2(b_1f_0)  + (p \cdot \partial)^2 (b_2f_0) \big], 
\end{equation}
and then take recourse to Grad's 14-moment approximation
\cite{Grad} for the single-particle distribution in orthogonal basis
\cite{Denicol:2010xn}.
This gives the desired equations:
\begin{eqnarray}
\Pi &=& \tilde a \Pi_{\rm NS} 
- \beta_{\dot \Pi} \tau_\Pi \dot \Pi
+ \tau_{\Pi n} n \cdot \dot u - l_{\Pi n} \partial \cdot n
- \delta_{\Pi\Pi} \Pi\theta
+ \lambda_{\Pi n} n \cdot \nabla \alpha
+ \lambda_{\Pi\pi} \pi_{\mu\nu} \sigma^{\mu\nu} \nonumber \\
&& + \Lambda_{\Pi\dot u} \dot u \cdot \dot u
+ \Lambda_{\Pi\omega} \omega_{\mu\nu} \omega^{\nu\mu} + (8 \ {\rm terms}) , \label{bulk}\\
n^\mu &=& \tilde a n^\mu_{\rm NS}
- \beta_{\dot n} \tau_n \dot n^{\langle \mu \rangle}
+ \lambda_{nn} n_\nu \omega^{\nu\mu}
- \delta_{nn} n^\mu \theta
+ l_{n \Pi}\nabla^\mu \Pi
- l_{n \pi}\Delta^{\mu\nu} \partial_\gamma \pi^\gamma_\nu
- \tau_{n \Pi} \Pi \dot u^\mu   \nonumber \\
&& - \tau_{n \pi}\pi^{\mu \nu} \dot u_\nu
+\lambda_{n\pi}n_\nu \pi^{\mu \nu}
+ \lambda_{n \Pi}\Pi n^\mu
+  \Lambda_{n \dot u} \omega^{\mu \nu} \dot u_\nu
+ \Lambda_{n \omega} \Delta^\mu_\nu \partial_\gamma \omega^{\gamma \nu}
+ (9 \ {\rm terms}), \label{heat}\\
\pi^{\mu\nu} &=& \tilde a \pi_{\rm NS}^{\mu\nu} 
-\beta_{\dot \pi} \tau_\pi \dot \pi^{\langle \mu\nu\rangle}
+ \tau_{\pi n} n^{\langle\mu}\dot u^{\nu\rangle}
+ l_{\pi n} \nabla^{\langle \mu}n^{\nu\rangle}
+ \lambda_{\pi\pi} \pi_\rho^{\langle \mu} \omega ^{\nu\rangle \rho}
- \lambda_{\pi n} n^{\langle\mu} \nabla^{\nu\rangle} \alpha
- \tau_{\pi\pi} \pi_\rho^{\langle\mu} \sigma^{\nu\rangle\rho} \nonumber \\
&& - \delta_{\pi\pi} \pi^{\mu\nu}\theta
+ \Lambda_{\pi\dot u} \dot u^{\langle \mu} \dot u^{\nu\rangle}
+ \Lambda_{\pi\omega} \omega_\rho^{\langle \mu} \omega^{\nu\rangle\rho}
+ \chi_1 \dot b_2 \pi^{\mu\nu}
+ \chi_2 \dot u^{\langle \mu} \nabla^{\nu\rangle} b_2
+ \chi_3 \nabla^{\langle \mu} \nabla^{\nu\rangle} b_2, \label{shear}
\end{eqnarray}
where ${\tilde a}=(1-a)$ and $\dot X = D X$.
The ``8 terms" (``9 terms'') involve second-order, linear
scalar (vector) combinations of derivatives of $b_1,b_2$.
All the terms in the above equations are inequivalent,
i.e., none can be expressed as a combination of others via equations
of motion
\cite{Bhattacharyya:2012nq}.
All the coefficients 
in Eqs. (\ref{bulk})-(\ref{shear})
can be written as functions of hydrodynamic variables 
\cite{Jaiswal:2012qm}.

In \cite{Jaiswal:2012qm}, 
we demonstrated the numerical significance of the new
dissipative equations derived here, by considering the evolution of a
massless Boltzmann gas, with the equation of state $\epsilon=3P$, at
vanishing net baryon number density, in the Bjorken model
\cite{Bjorken:1982qr}.

\section{Summary}

To summarize, we have presented a new derivation of the relativistic
dissipative hydrodynamic equations by introducing a nonlocal
generalization of the collision term in the Boltzmann equation. The
first-order (Navier-Stokes) and second-order (Israel-Stewart)
equations are modified: new terms occur and coefficients of others are
altered. While it is well known that the derivation based on the
generalized second law of thermodynamics misses some terms in the
second-order equations, we have shown that the standard derivation
based on kinetic theory also misses other terms. The method presented
here is able to generate all possible terms to a given order that are
allowed by symmetry.

\end{document}